# Diffusion-inspired time-varying phosphorescent decay in nanostructured environment

Denis Kislov[1], Denis Novitsky[1,2], Alexey Kadochkin[3], Dmitrii Redka[4] Alexander S. Shalin[1], and Pavel Ginzburg[5,6]

[1]ITMO University, 49 Kronverksky Pr., St. Petersburg 197101, Russia
[2]B.I. Stepanov Institute of Physics, National Academy of Sciences of Belarus, 68 Nezavisimosti Ave., Minsk 220072, Belarus
[3]Ulyanovsk State University, Ulyanovsk 432017, Russia
[4]Saint Petersburg
Electrotechnical University "LETI" (ETU) 5 Prof. Popova Street, St. Petersburg 197376, Russia
[5] Center for Photonics and 2D Materials, Moscow Institute of Physics and Technology, Dolgoprudny, 141700 Russia
[6]Tel Aviv University, Ramat Aviv, Tel Aviv 69978, Israel

**Abstract:**

Structured environment controls dynamics of light-matter interaction processes via modified local density of electromagnetic states. In typical scenarios, where nanosecond-scale fluorescent processes are involved, mechanical conformational changes of the environment during the interaction processes can be safely neglected. However, slow decaying phosphorescent complexes (e.g. lanthanides) can efficiently sense micro- and millisecond scale motion via near-field interactions. As the result, lifetime statistics can inherit information about nano-scale mechanical motion. Here we study light-matter interaction dynamics of phosphorescent dyes, diffusing in a proximity of a plasmonic nanoantenna. The interplay between the time-varying Purcell enhancement and stochastic motion of molecules is considered via a modified diffusion equation, and collective decay phenomena is analysed. Fluid properties, such as local temperature and diffusion coefficient are mapped on phosphorescent lifetime distribution extracted with the help of inverse Laplace transformation. The presented photonic platform enables contactless all-optical thermometry and diffusion measurements paving a way for a plethora of possible applications. In particular, the proposed analysis can be used for detailed studies of nanofluidic processes in lab-on-a-chip devices, which are extremely hard or even impossible to analyse with other optical methods.

*corresponding author:

denis.a.kislov@gmail.com

**Introduction**

Light-matter interaction processes strongly depend on both internal structure of quantum systems and a surrounding environment [1]. While the first factor in a vast majority of scenarios is pre-defined by nature, the latter on can be modified by carefully designed nanostructuring [2–6]. Purcell enhancement of spontaneous emission rates with plasmonic nanoantennae is one among prominent examples of tailoring first-order light-matter interaction, e.g. [7,8]. The essence of structuring-inspired manipulation can be understood via local density of electromagnetic states (LDOS), which enter interaction Hamiltonians governing quantum processes of any order [9,10]. LDOS is directly related to a classical Green's function, which allows formulating quantum problems in terms of pure classical electromagnetic quantities.

Common approach to solutions of problems involving light-emission processes is based on time-dependent perturbation theory in the case of weak coupling regime, or on the complete solution of dynamic equations in more advanced strong coupling scenarios. In both cases, however, strength of the fundamental interaction parameter is time independent. This approach is well-justified in cavity quantum electrodynamics, where the environment is static and does not change during the interaction. In colloidal applications, where a solution of particles can undergo either ordered or random motion, dynamical changes of the environment can be also neglected under certain and commonly met approximations. For example, fast virtual level assisted nonlinear processes and even dipole-allowed nano-second scale fluorescent decays are orders of magnitude faster than slow microsecond-scale changes in fluidic environment.

In a sharp contrast to dipole-allowed fluorescent transitions, phosphorescent dyes are characterized by micro- and milli- second slow decays, which originate from fundamentally different quantum process – spin flip followed by the photon emission transition. In particular, phosphorescent (e.g. [11]) and rare earth (e.g. [12]) luminescent molecules have characteristic decay times on the scale of micro to milliseconds. Those properties can be employed for different imaging techniques, such as gated photoluminescence microscopy [13] and also can be affected by nanostructuring[14–17]. In terms of fundamental light-matter interaction processes, phosphorescent timescales can be comparable with conformational changes inside colloidal environment. For example, a spherical molecule in water at room temperature has an average drift of several hundreds of nm during the time period of 100 μsec. This nano-scale displacement between an emitter and nanoantenna is enough to dramatically change the decay rate, as it will be shown hereafter. It is worth noting that conformational changes are still significantly slower than the optical carrier frequency and allow applying time-scale separation approach for the theoretical treatment. This approach has been recently applied for velocimetry mapping with phosphorescent emitters [18] and for analysis of rotational micro-Doppler effects [19,20].

Here we develop a theoretical framework for treating diffusion processes of slow-decaying phosphorescent compounds in a solution of resonant optical antennas (Fig. 1). An assembly of emitters, dissolved in a liquid mixed with metal nanoparticles, is pumped with an external illumination. Slow-decaying dyes diffuse in the vicinity of resonating nanoparticles, which change their emission on the timescale, comparable with the spontaneous decay. As the result of this interaction, the information on the fluid dynamics is imprinted in the photon statistics.

The manuscript is organized as follows: the basic diffusion model with an additional decay term associated with the position-dependent Purcell effect is introduced first. In contrary to traditional description of diffusion processes, the density of excited emitters is used as a variable. Orientation-averaged and position-dependent Purcell factor is calculated next. The analysis is based on the evaluation of classical Green's functions via Mie series expansion. Solution of the diffusion equation with the position-dependent Purcell-driven decay factor comes as the main result followed by the Conclusion.

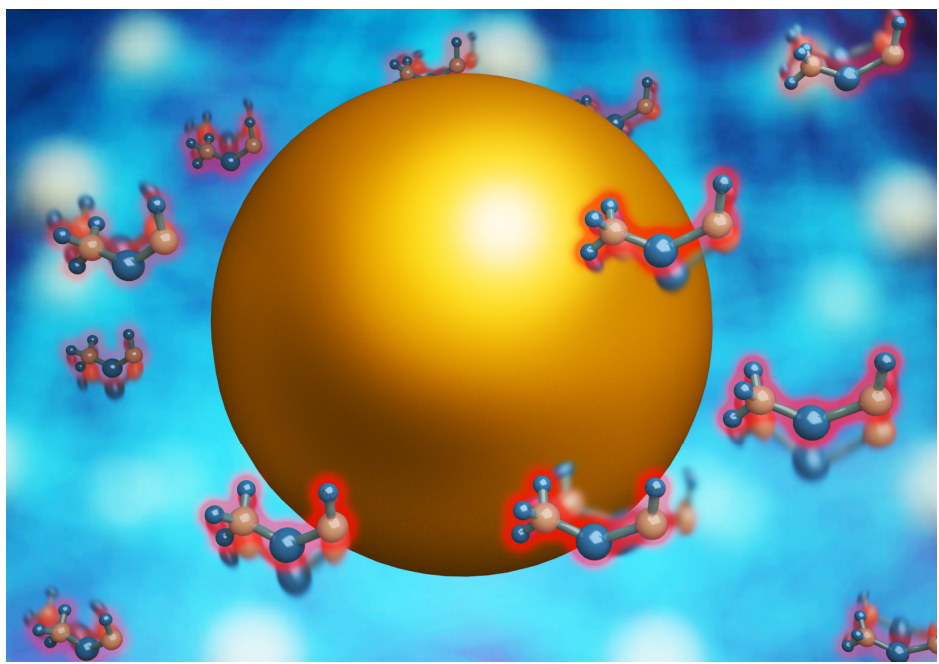

Fig. 1. The schematics of the system – diffusion of slow-decaying phosphorescent dyes next to a resonating nanoantenna.

**Theoretical Formalism**

*Diffusion Model*

The proposed diffusion model describes the density of excited and time-decaying molecules undergoing Brownian motion in a proximity of metal or dielectric nanoparticles (Fig. 1). Since the later objects have significantly higher mass, they can be assumed to be static during the interaction (on the other hand, the reference frame can be linked directly to the resonator, which is static in its rest frame). The concentration of excited molecules is $n(\mathbf{r},t)$ and it depends not only on the diffusion in a solvent, but also on time via Purcell-enhanced interactions with a nearby nanoantenna. This position-dependent decay rate can be written as $\gamma(\mathbf{r}) = \gamma_0 F(\mathbf{r})$, where $F(\mathbf{r})$ is the Purcell factor and $\mathbf{r}$ stays for the distance between the nanoparticle and the excited molecule. In free space, without a particle present, the characteristic decay time is $\tau_0 = 1/\gamma_0$. For lanthanide complexes it can vary from µs to ms depending on an internal structure and surrounding solvent, which can quench the radiation.

The diffusion equation for this type of a process can be written as:

$$\frac{\partial n}{\partial t} = D \Delta n - \gamma(\mathbf{r}) n \tag{1}$$

where *D* is the diffusion coefficient of phosphorescent molecules. Purcell effect introduces an additional spatially dependent decay channel. Diffusion characteristics (*D*) generally depend on temperature and other parameters of an environment, which can be related to each other via Stokes – Einstein relation:

$$\frac{D_{T_1}}{D_{T_2}} = \frac{T_1}{T_2} \frac{\mu_{T_2}}{\mu_{T_1}} \tag{2}$$

where *µ* is dynamic viscosity of the solvent and sub-indices correspond to different local temperatures [21]. This dependence can provide a new methodology for local temperature sensing via Purcell-effect-induced luminescence modification.

Hereinafter, the model will concentrate on the interaction dynamics with a single spherical particle by assuming radial symmetry of the entire scenario. It means that the pumping mechanism is also isotropic (unpolarized) and it induces a spatially dependent excitation of molecules in the region under consideration. While the antenna effect for pump can be also introduced, here for the sake of simplicity we assume a Gaussian initial distribution of the form $n(r,0) = n_0\, e^{-(r-a)^2/\Delta r^2}$, where *a* is the particle's radius and $\Delta r$ the effective spot size of tightly focused pump laser beam. Under the simplifying conditions, we can neglect the angular dependence of concentration and recast Eq. (1) in spherical coordinates by leaving only the radial component:

$$\frac{\partial n}{\partial t} = D\frac{\partial^2 n}{\partial r^2} + 2\frac{D}{r}\frac{\partial n}{\partial r} - \gamma(r)n \tag{3}$$

We will solve Eq. (3) numerically in the range of *r* from the radius of the nanoparticle *a* to a certain value *R>>a*. This allows us to set the boundary conditions as an absence of the molecular flow near the surface and far away from the nanoparticle $\frac{\partial n(0,t)}{\partial r} = \frac{\partial n(R,t)}{\partial r} = 0$. This assumption relies on relatively low density of nanoparticles in the liquid. In order to obtain the solution of Eq. 3, position-dependent Purcell Factor should be estimated, which will be done next.

*Position and orientation averaged Purcell enhancement*

Electromagnetic Green's function of a radiating dipole next to spherical dielectric particle has a closed form analytical solution based on Mie theory. This allows to obtain the analytical expressions for Purcell factor describing the enhancement of spontaneous emission rates in the vicinity of nanoparticles as follows [22,23]:

$$F_\perp^{rad}(r) = \frac{3}{2}\sum_{m=1}^{\infty} m(m+1)(2m+1)\left|\frac{\psi_m(kr\sqrt{\varepsilon_d})}{(kr\sqrt{\varepsilon_d})^2} + A_m\frac{\xi_m(kr\sqrt{\varepsilon_d})}{(kr\sqrt{\varepsilon_d})^2}\right|^2$$

$$F_\parallel^{rad}(r) = \frac{3}{4}\sum_{m=1}^{\infty}(2m+1)\left(\left|\frac{\psi_m(kr\sqrt{\varepsilon_d})}{kr\sqrt{\varepsilon_d}} + B_m\frac{\xi_m(kr\sqrt{\varepsilon_d})}{kr\sqrt{\varepsilon_d}}\right|^2 + \left|\frac{\psi'_m(kr\sqrt{\varepsilon_d})}{kr\sqrt{\varepsilon_d}} + A_m\frac{\xi'_m(kr\sqrt{\varepsilon_d})}{kr\sqrt{\varepsilon_d}}\right|^2\right), \tag{4}$$

where the values stay for radiative enhancement of the emission in the case, when the dipole moment of a molecule is either perpendicular or parallel to the nanosphere surface. In the case, when the molecules are randomly oriented in respect to the sphere, the average enhancement is given as $F^{rad}(r) = \frac{2}{3}F_\parallel^{rad}(r) + \frac{1}{3}F_\perp^{rad}(r)$. Corresponding total rate enhancements are given by

$$F_\parallel^{tot}(r) = 1 + \frac{3}{4}\sum_{m=1}^{\infty}(2m+1)\text{Re}\left[B_m\left(\frac{\xi_m(kr\sqrt{\varepsilon_d})}{kr\sqrt{\varepsilon_d}}\right)^2 + A_m\left(\frac{\xi'_m(kr\sqrt{\varepsilon_d})}{kr\sqrt{\varepsilon_d}}\right)^2\right],$$

$$F_\perp^{tot}(r) = 1 + \frac{3}{2}\sum_{m=1}^{\infty} m(m+1)(2m+1)\text{Re}\left[A_m\left(\frac{\xi_m(kr\sqrt{\varepsilon_d})}{(kr\sqrt{\varepsilon_d})^2}\right)^2\right]. \tag{5}$$

In Eqs. 4 and 5 $\varepsilon_d$ is the dielectric permittivity of the environment, *k* - wavenumber in vacuum, $A_m$ and $B_m$ are Mie coefficients (widely used notions have been used [22,23]), *m* – number of multipole terms,

$\psi_m(x) = x j_m(x)$, $\xi_m(x) = x h_m^{(1)}(x)$ where $j_m(x)$ and $h_m^{(1)}(x)$ spherical Bessel and Hankel functions of the first kind.

Radiative and nonradiative enhancements should be distinguished, as they influence the light-matter interaction in different ways. Total radiative lifetime governs the decay dynamics, while the radiative contribution is responsible for the number of photons detected at the far field. The difference between radiative and total rates is the result of losses within the particle[24]. While in a majority of optoelectronic applications only the radiative enhancement is a factor for maximization, our diffusion model requires the knowledge of the total decay rate. The information about local properties of the fluid, however, can be analyzed via collecting emitted photons. Separation into radiative and nonradiative channels based on Green's functions formalism can be also preformed and will be discussed in later sections [25].

Radiative and total enhancements are plotted as the function of the distance from the gold nanoparticle (Fig. 2). The following typical parameters have been used – the radius of the spherical particle is 50 nm, the optical properties of gold were taken from [26]. The phosphorescent emission central wavelength is 690 nm. It can be seen that nonradiative channels prevail the decay dynamics in the close proximity of the particle, while at distances larger than 100nm the influence of the particle is minor, and Purcell enhancement approaches unity (no enhancement). Qualitatively, it means that far-situated fluorophores will contribute to the background radiation and will decay with the rate of $\gamma_0$. Those contributions can be factorized by applying lifetime distribution post-processing techniques [27]. The most relevant region for observing the diffusion dynamics via photon counting is situated at distances of 20-100 nm from the nanoparticle surface.

*Temperature dependence of the diffusion coefficient*

Taking into account the spatial distribution of the Purcell enhancement and low density of the dissolved nanoparticles, electromagnetic coupling between neighboring nanoantennas can be ignored. Furthermore, contributions of the emitters situated far apart from the particle are neglected, since their lifetimes are unmodified during the diffusion process, as it was previously discussed.

The diffusion coefficient of emitters is given by the Stokes-Einstein relation assuming spherical shapes of the molecules and low Reynolds numbers, which are justified in the case of nano-scale objects dissolved in water[28]:

$$D(T) = \frac{k_B T}{6\pi \cdot \mu(T) \cdot R_{mol}}, \quad (6)$$

where $\mu$ is the dynamical viscosity of the solvent, $T$ - absolute temperature, $R_{mol}$ - the radius of the light-emitting complex. $R_{mol} \approx 0.5-1 nm$ for the molecules, which are discussed here[28]. There are empirical models relating dynamical viscosity and local temperature. For example, distilled water obeys $\mu(T) = A_0 \cdot 10^{B_0/{T-C_0}}$, where $A_0 = 2.4 \cdot 10^{-5} Pa \cdot s$, $B_0 = 247.8K$ и $C_0 = 140K$. In this case, the diffusion coefficient for a range of organic compounds, e.g. (eosin, rose bengal, erythrosine), which possess relevant phosphorescent properties [29], will fall in the following range: $0.2 \div 1.6 \, \mu m^2/ms$ (from 273 to 373 K).

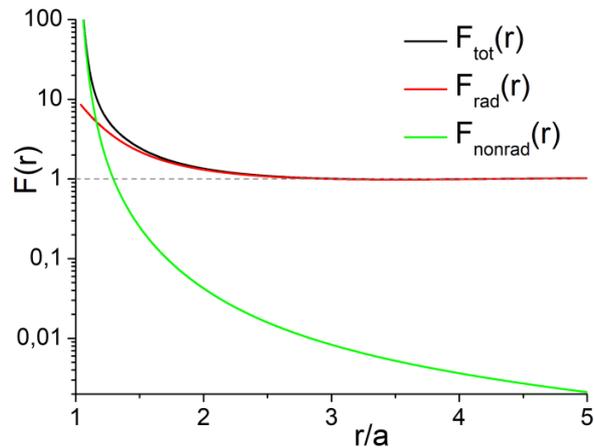

Fig. 2. Purcell enhancement next to a gold (50nm radius) nanoparticle. Orientation-averaged total, radiative and nonradiative enhancements as the function of the normalized distance (to the particle's radius) between the dipole and particle's surface appear as black, red and green lines respectively. Detailed parameters are given in the main text.

*Analysis of the diffusion-inspired emission dynamics*

Having the values of the diffusion coefficient and the position-dependent Purcell factor, the entire model can be solved now. Fig. 3 shows the density of the excited molecules as the function of the distance from the particle at different instances of time. Color lines correspond to different times elapsed from the instance of the pulsed pump excitation.

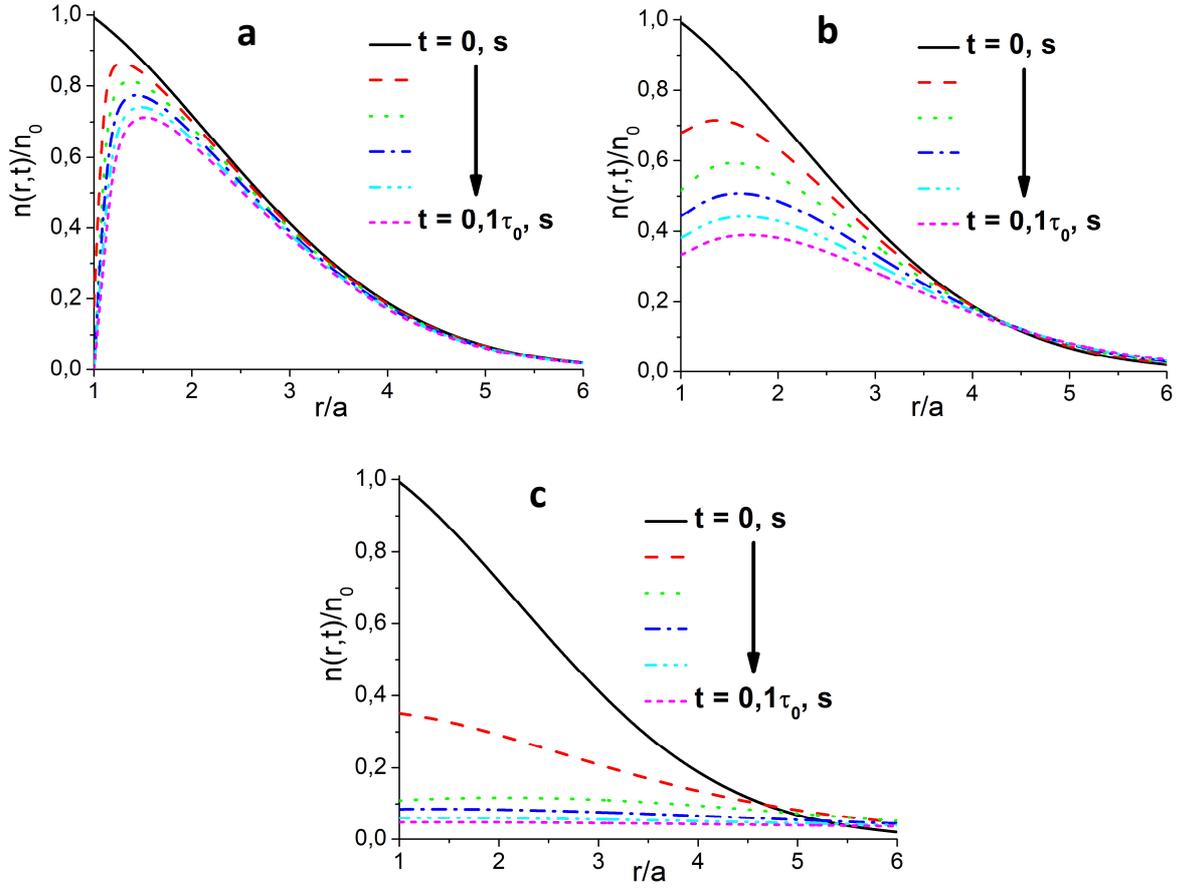

Fig. 3. Radial distribution of excited dye molecules in a vicinity of the particle at different times for different diffusion coefficient values: a) D=0, b) 0.2, c) 1.6, $\mu m^2/ms$. Parameters used in the calculations: $a = 50 nm$, $\Delta r = 3.4a$, $\tau_0 = 300 \mu s$.

It can be seen from Fig. 3 that when the diffusion coefficient is very small (panel a) the population of excited dyes drops down fast next to the particle and there is no Brownian inflow towards it. On the other hand, when the diffusion is efficient, slow decaying molecules (unaffected by Purcell enhancement) flow in and start sensing the presence of the antenna. As the result, they become decaying faster. This dynamical behavior is clearly seen by comparing panels a, b, and c. The decay kinetics has a direct replica on the lifetime distribution, which can be measured at the far-field.

Intensity collected at the far-field has the following time dependence:

$$I(t) \sim \int_a^{R_{collection}} \int_0^{\pi} \int_0^{2\pi} F^{rad}(r) \gamma_0 n(r,t) r^2 \sin^2(\vartheta) dr d\vartheta d\varphi. \quad (7)$$

$R_{collection}$ here corresponds to the focal volume of a collecting objective; it is supposed to be much larger than the spot size of the exciting beam $\Delta r$ (this assumption does not affect the main result, however).

Note that the time dependence solely originates from $n(r,t)$. While $n(r,t)$ is governed by the total decay rate, the emitted intensity in Eq. (7) also depends explicitly on the radiative Purcell enhancement. Fig. 4(a) demonstrates the time-dependent intensity decay for different values of the diffusion constant. It is clearly seen that in the chosen range of parameters the intensity drops faster with the increase of the diffusivity. When the molecules are randomly moving around the antenna, they have larger probability to be found in its vicinity and, as a result, they experience larger Purcell enhancement.

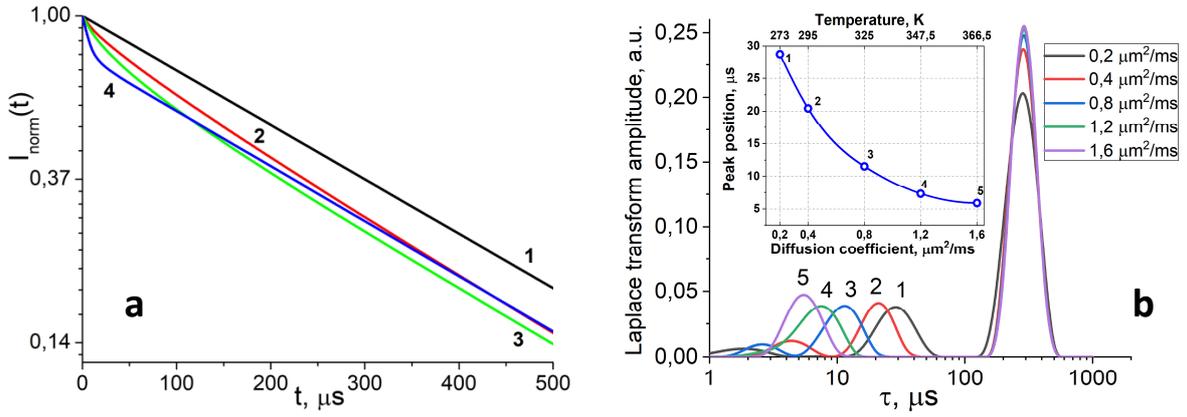

Fig. 4. (a) Normalized intensity (log-scale) decay of the dye molecules in a vicinity of the particle for different diffusion coefficient values: 1 - Intrinsic phosphorescence of erythrosine molecules in water; 2 - D=0; 3 - $0.2$; 4 - $1.6$, $\mu m^2/ms$. (b) Lifetime distribution analysis of the collected luminescence signal. Color lines correspond to different diffusion coefficients and, hence, temperatures. The inset shows the position of the secondary (numbered) peak as the function of the diffusion coefficient. The decay time value is derived from the regularized inverse Laplace transformation. The radius of the collecting objective focal volume is $R_{collection}=10a$.

The kinetics of the collected intensity can be used as a tool for diffusion and, hence, temperature detection. To demonstrate this, we apply the inverse Laplace transformation on the function $I(t)$ given by Eq. (7). Namely, the intensity can be represented as

$$I(t)=\int_0^\infty g(s)e^{-st}ds, \qquad (8)$$

where $s=1/\tau$ is the inverse relaxation time. We solve this equation using the numerical approach similar to that reported in Refs.[27] and get the distribution $g(s)$ of different exponential contributions to the total luminescence kinetics. The result is shown in Fig. 4(b), where the nonexponential response of the system is clearly seen. The right strongest peak corresponds to the free-space relaxation time of the molecules ($\tau_0=300\mu s$). The secondary peak on the left is associated to the Purcell effect and contains the information on the diffusion characteristics. Indeed, it is seen that the increase in the diffusion coefficient results in further shift of this secondary peak towards shorter lifetimes due to increasing probability for molecules to approach to the metal nanoparticle. The position of this peak as a function of the diffusion coefficient and, consequently - temperature is shown in the inset. Therefore, it becomes possible to measure temperature and diffusion coefficient in a liquid via the proposed photonic contactless design. This pronounced dependence of the optical signature of phosphorescent molecules on surroundings can be extremely useful for, e.g., microfluidics and microchamber-based chemistry, remote all-optical temperature control, and microbiology and biomedicine to name just few.

**Conclusions**

Lab-on-a-chip concept is a very fast-growing field attracting ever-increasing attention, as it opens a room of opportunities for a plenty of highly demanded applications. However, hardware realizations of many important functionalities are often face challenges, especially in cases when nano-scale resolution is required. Here we developed a novel concept for contactless all-optical temperature and diffusion measurements enabled by dynamic time-dependent Purcell effect in a solution of phosphorescent molecules interfacing resonant nanoantennae. Dynamics of the long life-time phosphorescent molecules decay is shown to be strongly dependent on the Brownian motion next to a resonator. This special interaction is described with temperature and diffusion coefficient of the surrounding liquid. Subsequently, far-field radiation emitted from diffusing molecules is analyzed via the inverse Laplace transform and exploited to recover local properties of a fluid environment. As a result, an efficient contact-free approach to measure required hydrodynamical characteristics of a liquid in a broad temperature range (e.g. water at the range of 273 to 373 K or any other fluid) and with nano-scale spatial resolution is demonstrated. Moreover, the proposed method can utilize biologically compatible compounds demonstrating new capabilities in a variety of lab-on-a-chip realizations and expanding the range of microfluidics applications.

**Acknowledgments**

The work has been supported in part by ERC StG 'In Motion'. A.S.S acknowledges the support of the Russian Fund for Basic Research within the projects 18-02-00414, 18-52-00005. Numerical simulations of the particles dynamics and lifetimes have been supported by the Russian Science Foundation (Project No. 18-72-10127).

**Conflict of Interest**

The authors declare no conflict of interest.